\documentclass[submission,copyright,creativecommons]{eptcs}
\usepackage{graphicx}
\usepackage{float}
\title{RTL2RTL Formal Equivalence : Boosting the Design Confidence}
\author{M V Achutha KiranKumar
\institute{Intel Corporation}
\email{achutha.kirankumar.v.m@intel.com}
\and
Aarti Gupta
\institute{Intel Corporation}
\email{\quad aarti.gupta@intel.com }
\and
S S Bindumadhava
\institute{Intel Corporation}
\email{\quad bindumadhava.ss@intel.com }
}
\def\titlerunning{RTL2RTL Formal Equivalence}
\def\authorrunning{KiranKumar, Aarti \& Bindumadhava}
\begin{document}

\maketitle

\begin{abstract}
Increasing design complexity driven by feature and performance requirements and the Time to Market (TTM) constraints force a faster design and
validation closure. This in turn enforces novel ways of identifying and debugging behavioral inconsistencies early in the design cycle. Addition of incremental
features and timing fixes may alter the legacy design behavior and would inadvertently result in undesirable bugs. The most common method of
verifying the correctness of the changed design is to run a dynamic regression test suite before and after the intended changes and compare the results, a method which is not exhaustive.
Modern Formal Verification (FV) techniques involving new methods of proving Sequential
Hardware Equivalence enabled a new set of solutions for the given problem, with complete coverage guarantee. Formal Equivalence can be applied for proving functional integrity after design changes resulting from a wide variety of reasons, ranging from
simple pipeline optimizations to complex logic redistributions. We present here our experience of successfully applying the RTL to RTL
(RTL2RTL) Formal Verification across a wide spectrum of problems on a Graphics design. The RTL2RTL FV enabled checking the design sanity in a very short time, thus
enabling faster and safer design churn. The techniques presented in this paper are applicable to any complex hardware design.

\end{abstract}
\section{Introduction}
\label{intro}
Graphics designs are increasingly finding their relevance in
new market segments like smartphones, tablets and a faster
churn of optimized designs is most desirable to meet the market requirements. Verifying the
design for incremental changes is an involved challenge and is the most
time consuming and critical aspect of the design process.
Traditional DV methods mandate a compromise between
breadth of coverage and resources available. We have
pioneered methodologies to enable formal verification at
the design stage which help a faster churn of RTL and early
stabilization.\par 
Formal verification has proved to be an ideal candidate to
verify tough SoC design challenges due to its ability to
exhaustively verify all possible complex scenarios without
any need for a test bench or input stimulus. With formal, a
designer or verification engineer need not spend
time stimulating all possible scenarios as the formal
engines carry out this task under the hood and result in
increased confidence. This eliminates the uncertainty of not
verifying a scenario which is either difficult to think of or is
missed due to the complexity of the design.\par 

Formal equivalence is a known field of research and the
most common application of the methodology is in
checking the correctness of the netlist generated by the
design synthesis against the RTL which is synthesized. The
ability to formally determine functional equivalence
between RTL models is a key enabler in physically aware
front-end design methodologies that are being practiced in
high performance designs. Earlier Combinational
Equivalence checking tools needed state matching designs
which were tested with equivalent functional maps. With the
advent of new methodologies and sequential equivalence
tools where the sequentially different implementations of
the designs could be verified for functional equivalence, a lot of problems which were
tough to be checked earlier, came under the gamut
of formal verification and hence the faster verification of
the design changes retaining the legacy behavior became a
possibility.\par 

Combinational equivalence checking (CEC) plays an
important role in EDA \cite{paper12,paper11,paper10}. Its immediate application is
verifying functional equivalence of combinational circuits
after multi-level logic synthesis \cite{paper7}. In a typical scenario,
there are two structurally different implementations of the
same design, and the problem is to prove their functional
equivalence. This problem was addressed in numerous
research publications. But the CEC could not solve the
problems when the logic was moved across the equivalent
states and selective disabling of the check of those unequal
states was an involved process. CEC had limitations when
the designs to be compared were not state matching. There
were some clever techniques applied to resolve individual
problems but a comprehensive solution needed a check
beyond CEC. With the increasing use of sequential
optimizations during logic synthesis, sequential
equivalence checking (SEC) \cite{paper1} has become an important
practical verification problem. SEC might employ symbolic
algorithms, based on binary decision diagrams (BDD) to
traverse the state space or any optimized methodology for
the specific state space traversal to check for the
equivalence of two circuits. On the other hand, the
equivalence problem could also be mapped to a model
checking problem, where a set of properties 
define the equivalence between the two circuits. Instead of a
set of properties, the formal sequential equivalence
checking (SEC) may adopt a reference model (RM), which
is a description at a higher level abstraction of the
functionality.\par 

As discussed, the SEC can be a check of RTL against the
high level reference model or the RM can be another piece
of RTL itself. The scope of discussion of this paper is limited to the equivalency check between two different
RTL models.\par

The organization of this paper is as follows: Section 2
gives a detailed explanation of all experimented areas of
application. Section 3 details the complexity reduction
techniques tried during the equivalence checking and some
best practices that have been deployed successfully. Section
4 talks about some more applications where there could
be a definite potential of applying SEC and the paper
concludes in Section 5 detailing some of the results
achieved by employing the RTL2RTL FV.
\section{Areas of Application}
\subsection{Parameterization}
\begin{figure}[H]
 \begin{center}
 \includegraphics [scale = 0.9] {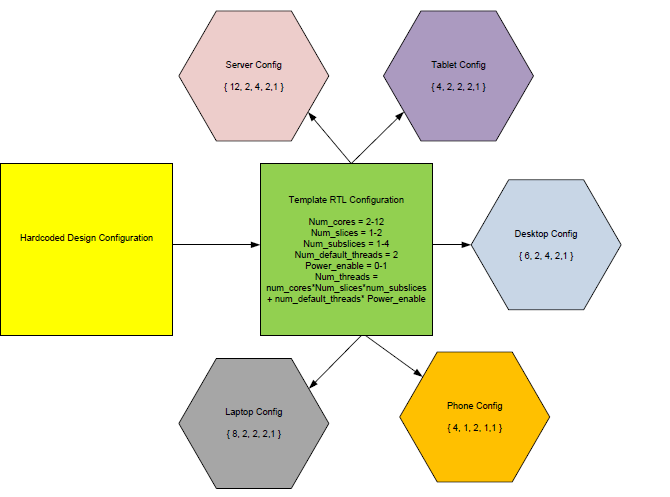}
 \caption{\small Avatars possible through Parametrization}
 \label{parameterization}
 \end{center}
 \end{figure}

Parameterization is the process of deciding and defining the
parameters necessary for a complete or relevant
specification of a design. Most of the legacy designs do
start with hard coding the parameters for a design and as
requirements press on the usage of the design in various
configurations, the design team resorts to parameterize
some of the common parameters. Figure \ref{parameterization} talks about the
many proliferations of one of the graphics design where it
could be targeted to a wide variety of markets.\par

As mentioned in the section \ref{intro}, the common form of proving
the correctness of the design is running an existing dynamic
regression over a selected set of seeds and number, just to
guarantee that the design has not been compromised on the
legacy behavior. But Validation is never complete and
comprehensive and there were some corner cases which are
always exposed by just running the DV. An obvious
verification for this kind of problem is to run FV on the
design with the non-parameterized code as Specification
model (SPEC) and the parameterized code with the default
parameters as the implementation model (IMP).

A positive validation for the base parameters against a non-
parameterized code is only possible with such equivalence.
For the other valid parameter settings, other dynamic /
formal validation techniques would be needed to assess the
correct programming. One interesting scenario was
observed when a negative formal equivalence was tried
with a non-default parameter setting. This kind of
equivalency check could be handled by CEC and SEC
tools.
\subsection{Timing Fixes - Logic Redistribution}

Fixing Critical timing paths is one of the common activities
for any synchronous design. One of the common solutions
for fixing critical paths is to redistribute the logic across the
different pipeline stages after resorting to all optimizations\cite{paper2}.
Studies suggest that a decent amount of unintentional functional bugs are
introduced while fixing the timing issues. These kinds of failure scenarios could be
easily avoided by running formal verification making sure the
design retains its sanity irrespective of the logic
redistribution across pipelines.
\newline
\begin{figure}[H]
 \begin{center}
 \includegraphics [scale = 0.7] {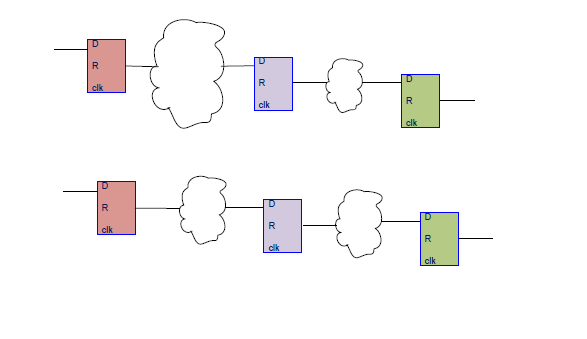}
 \caption{\small Logic Redistribution for Timing Violation Fix}
 \label{timing}
 \end{center}
 \end{figure}
As shown in Figure \ref{timing}, there is a huge combinational path in
between the first and second flops which would be
redistributed to fix the violation. In this kind of case, a
combinational equivalence checker would fail, as the states
would not match across the designs and the sequential
equivalence checker is the appropriate solution.
\subsection{Timing Fixes - Critical Path Reduction}

\begin{figure}[H]
 \begin{center}
 \includegraphics [scale = 0.8] {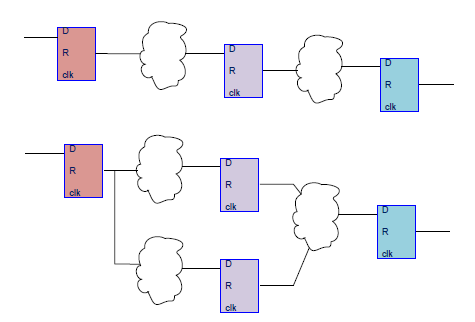}
 \caption{\small Critical Path Reduction for Timing Fixes}
 \label{critical path}
 \end{center}
 \end{figure}

Where certain timing fixes would not have the flexibility to
redistribute the logic across pipe stages, the designers 
bifurcate the computation logic and redirect some of the
computation logic through a parallel flop path as shown in
Figure \ref{critical path}.
\newline
As there is a new parallel pipeline path, this would not be a
straight forward problem for the combinational checker to
solve. Some tools allow skipping one identified stage of
checking but for convergence reasons, mandate the check
at the following pipeline stage. While some commercial
tools would be intelligently handling these kinds of cases,
sometimes, it would need a user intervention for
convergence reasons. SEC would be more ideal in these
kind of scenarios compared to CEC from proof convergence perspective.
\subsection{Pipeline Optimizations}

As manufacturing processes, algorithms and
implementation methodologies mature over time,
computing pipeline depth gets optimized and the functionality could be computed within a reduced pipeline depth. On the
contrary, some of the timing fixes and other algorithm
requirements may require an addition of extra
pipeline stage while retaining the functionality.

\begin{figure}[H]
 \begin{center}
 \includegraphics [scale = 0.5] {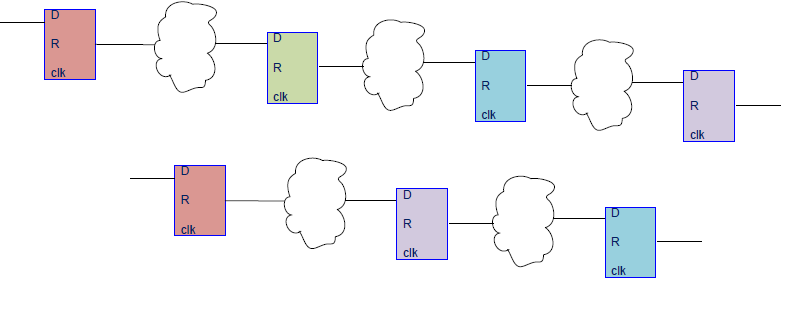}
 \caption{\small Pipeline Optimization}
 \label{Pipeline}
 \end{center}
 \end{figure}
Any of the afore mentioned cases, pipeline optimizations or timing fixes, would need additional states defined in the design. This would need an equivalence check on non-state matching designs. 
Depending on the case of re-pipelining optimization, one of the above
depicted design portion can be considered as SPEC (or the specification/golden reference) and the other IMP (or the implementation/changed design).
But the end to end functionality remains unperturbed
and hence formal verification would be the ideal solution to
guarantee that the functionality is preserved. SEC becomes the methodology of choice for non-state matching design equivalence.
The equivalence check would necessitate defining the
latencies for both the specification and
implementation models.

\subsection{Chicken Bit Validation}

Chicken bits are the bits exposed to the driver to disable a
feature in silicon. It is intended to revert the design changes made where the confidence is not high (confidence is directly
proportional to validation efforts). It is next to impossible
to hit every possible state in pre-silicon through DV. Most of the design
fixes these days do implement chicken bits and many times,
these chicken bits unintentionally affect the real
functionality. Typically most of the critical features account
for chicken bits early in the design cycle. But there would
be a small fraction of the number of chicken bits that are
added towards the end of the design cycle, to give the
flexibility to disable those diffident features. Negative
validation of these kind of chicken bits is 
challenging, as a feature disabling is as
intrusive in the code as the feature itself.
\newline
\begin{figure}[htbp]
 \begin{center}
 \includegraphics [scale = 0.8] {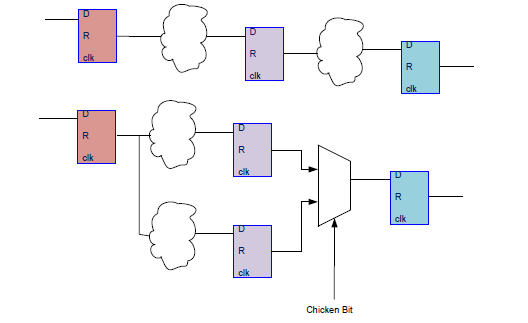}
 \caption{\small Chicken bit Added in the design}
 \label{chknbit}
 \end{center}
 \end{figure}
Most of the designs do implement the chicken bit in the
mode depicted in Figure \ref{chknbit}. One way to guarantee
the correctness of the design changes is to run formal
equivalence verification of the earlier design against the
current design in question with the chicken bit disabled.
Though this sort of verification does not need a sequential
logic checker, a good debugging capability of the tool used
would ease the life of the designer to fix any issues
reported.
\subsection{Clock gating Verification}

One of the most commonly used low-power techniques in
any design is clock gating and proper clock gating is
necessary for data integrity. If clock is not shut down
properly for a piece of logic, improper state signals and
signal glitches can propagate and lead to data corruption.
\par
The default verification strategy for a clock-gated design is
to run the set of golden regression test suite on both the pre-clock-gated design and the post-clock-gated design, with the
assumption that the golden test suite exercises all corner-cases when clocks would be gated. However, this assumption is not
always true, especially in less aggressive clock-gating
schemes. Coverage of the corner cases is always
challenging as the existing suite might not expose all scenarios. The best strategy is to use Sequential
Equivalence Checking tools for RTL vs. RTL comparison.
CEC might not be an obvious preference for this kind of design check.\par
\begin{figure}[H]
 \begin{center}
 \includegraphics [scale = 0.8] {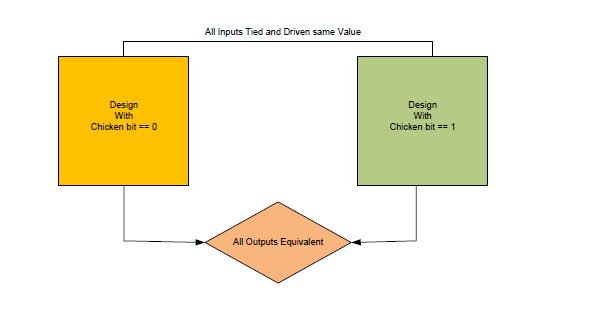}
 \caption{\small Clock Gating Verification}
 \label{clkgate}
 \end{center}
 \end{figure}

As depicted in the Figure \ref{clkgate}, FEV is run on a design with
clock gating enabled on IMP and disabled in SPEC, to
make sure that the design behaves exactly similar in all
scenarios. Some of the cases would need addition of
realistic constraints on the inputs. These
constraints would assist in verifying the real intent and
hence assist in convergence. 

\subsection{Power Aware Equivalence Verification}

Low power design continues to garner increased attention
in nearly every design segment. Many design
techniques have been developed to reduce power and by
the judicious application of these techniques, systems are
tuned for the best power/performance trade-offs. Design
sign-off with great quality is essential to avoid re-spins
while meeting market pressure. Low power specification
defined in UPF (Unified Power Format) introduces certain
power-logic such as, insertion of isolation cells or retention
mapping during synthesis.\par 
With functional intent being separated from power-intent,
the need for power-aware logical equivalence check (EC)
methodology is indispensable. Formal equivalence of
designs with and without UPF could result in checking if
the power intent (UPF) is syntactically correct and checks
for incorrect/missing/inconsistent isolation rules.
\newline
\begin{figure}[H]
 \begin{center}
 \includegraphics [scale = 0.9] {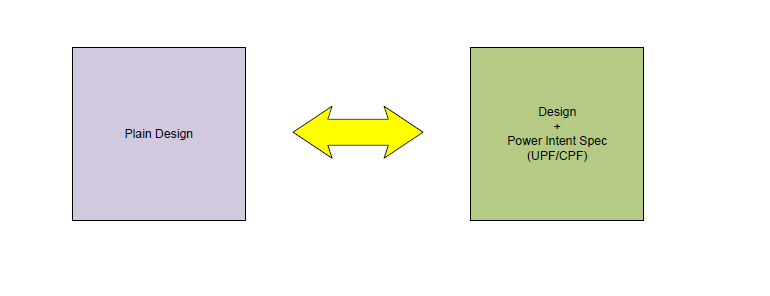}
 \caption{\small Power Intent Equivalence Verification}
 \label{Power}
 \end{center}
 \end{figure}
There are various tools available in both CEC and SEC
which could handle such equivalence including the power
intent. A judicious selection of the tool which can help
faster debug and convergence is advised. Any low power
tool in industry can do these kind of checks.
\subsection{Basic X-Checking: Uninitialized Flops}

Almost all designs would have state elements in which
some of them would be initialized to a defined value post
reset and some which are not. Out of reset, these
uninitialized state elements can come out in any state on the
silicon and to represent the same, in simulation, most of the
non-2 valued simulators would bring out those elements as
``X''s. Formal Verification would also bring out those
uninitialized state elements in undefined state which can
take any value of 0/1. All RTL2RTL formal verification
tools provide a utility to define the behavior of all flops
inside the design that are not connected to reset. Our
methodology of checking on those types of Xs by first
initializing those elements to be 0s and make sure the
equivalence test passes, followed by removal of such
constraint which would bring out the Xs emanating from
those flops and the comparison at the output would result in
counter example for the same design used as SPEC and
IMP.
\newline
\begin{figure}[H]
 \begin{center}
 \includegraphics [scale = 0.85] {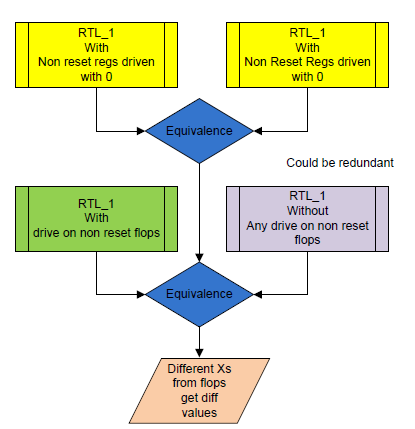}
 \caption{\small X checking from uninitialized flops}
 \label{uninitflops}
 \end{center}
 \end{figure}
An X emanating from SPEC would not be equal to the X
from IMP and hence the counter example would point to
the differing state assignment. While deriving the Xs
because of the flops, the design could be elaborated where
the undriven nets and specific X assignments can be
considered as symbolic.\par

An alternate mode of checking that some flops are driving unwanted values that propagate to the outputs is by assigning all
non-resettable state elements to 0 in one design (SPEC) and
1 in the other (IMP). This method of checking is not
completely comprehensive, as we might miss some of the critical
combinations of 0s and 1s of different state elements. But this mode of checking can
converge faster and can catch all the low hanging fruits
faster and is useful in bug hunting mode.
\subsection{Basic X Checks: Xs from undriven nets or internal assigned Xs or Stopats:}

Xs in design can be due to various reasons other than just
from the uninitialized state elements. Similar to the above X checking
methodology described in section 2.8, the Xs from the direct
assignments and the un-driven nets could be derived from
the design. The design elaborated with X assumed to be 0/1 and the design elaborated
without any such assumption are compared to check for the propogation of Xs.\par 

\begin{figure}[htbp]
 \begin{center}
 \includegraphics [scale = 0.9] {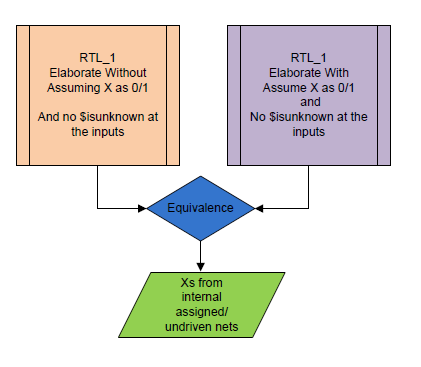}
 \caption{\small X-checking on Internal X-assignments or Undriven Nets}
 \label{intundriven}
 \end{center}
 \end{figure}

As a prelude to the X checking due to undriven or internal 
assignments, the tools can be tuned to assume the non-resettable flop elements come out with an equivalent 
definitive value in both the designs. This methodology 
would still avoid the checking of X’s due to out of bound 
  array interactions. A tool capable of X-controlling 
  capability can handle this kind of checks. As mentioned in 
  the subsection above, this methodology is very helpful in 
  the bug hunting mode. The constraints are absolutely 
  needed to thwart out Xs being reported from non-functional 
  cases. There had been a lot of cases where we experienced 
  bogus Xs because the inputs are not constrained for the 
  valid set of inputs and hence Xs for those vectors are truly 
  ignorable. 
\subsection{New Feature Addition with Backward Compatibility}

There are certain designs which are very critical to the
product, as the operations of those units are exposed to the
external customer directly and the design has to stick to
some standards of functionality. As an example, the
execution unit of the design needs to stick to some standard
(IEEE, DX*, OCL, OGL, etc...). The operations handled
by the Execution unit will be the instruction set for the
design.\par 

\begin{figure}[H]
 \begin{center}
 \includegraphics [scale = 0.5] {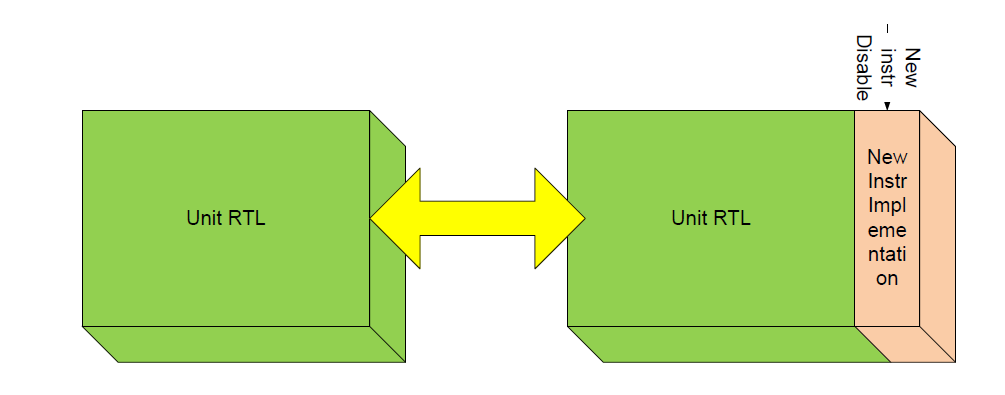}
 \caption{\small Confirming Legacy behaviors with feature addition}
 \label{legacy}
 \end{center}
 \end{figure}

Most of such designs would have reverse compatibility
w.r.t the previous generations of the design, which would
mean that the instructions which were implemented in the
earlier generation would retain the same functionality and
the new feature or the instruction implemented should not
have tampered with all other features or instructions
implemented. This is not applicable to optimizations or
changes in the specified behavior of the previous
instructions/features.\par

Most of our new feature additions or opcode additions are
taken through the reverse compatibility analysis where the
inputs are constrained for disabling the new feature/opcode
being added and is checked against the legacy design. This
has helped us in maintaining the legacy behavior in spite of
new optimization/feature/opcode additions.
\subsection{Replacing a Big Data regression with FEV}
\label{bigdata_section}

Most of the data path operations are taken through STE \cite{paper9}
regression for the implemented proofs at Intel\textsuperscript{\textregistered}. Proving
each and every operation for every model would take
anywhere between 3-5 days based on the net batch resource
availability and machine configuration.
\newline
\begin{figure}[H]
 \begin{center}
 \includegraphics [scale = 0.7] {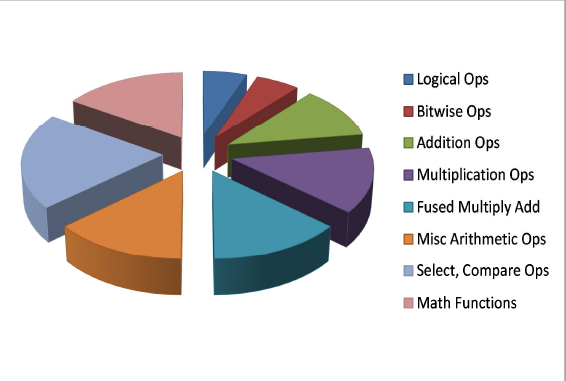}
 \caption{\small Operations divided into buckets for equivalence}
 \label{bigdata}
 \end{center}
 \end{figure}
With the advent of powerful equivalence tools, one of the
optimization made in the recent times is to run the regression against the previous
STE proven implementation. In order to assist in faster convergence, the
operations are subdivided into different buckets and the
whole regression completes in less than 1 hour on one
machine. Comparatively, the complete STE regression used
to take 10 machines and decent memory on these resources.
The figure \ref{bigdata} depicts the percentage of the time
taken by different sets of operations in the whole regression
time.
\subsection{Regular FPV}

One of the most popular techniques in formal property
verification is to write out an abstract model of the design
similar to RTL and write properties to define the
equivalence of the two design outputs. The same
methodology can be easily verified through equivalence
where the specification is the abstract model and
implementation is the real piece of code in the RTL. The
regular model checking engines are not as optimized as the
equivalence tool engines and the convergence of such
checks had always been challenging. Most of such abstract
models written might not match the states with the real
implemented design and hence the CEC wouldn’t be an
ideal choice for such comparison.

\section{Complexity Reduction Techniques}
Like any formal verification, RTL2RTL FEV also has
capacity limitations. A range of techniques are applied to
overcome the capacity issues. This section discusses in
detail about such techniques used to solve the complexity
of the FV task.
\subsection{Divide and Conquer Approach}

To overcome the capacity issue, typically FV is handled
using a divide and conquer approach, similar to the
compositional verification \cite{paper3}. A typical compositional
approach would be to decompose the whole problem into
number of sub proof tasks and prove each of them
independently and rework at the top level with these sub
blocks black boxed.\par

\begin{figure}[H]
 \begin{center}
 \includegraphics [scale = 0.85] {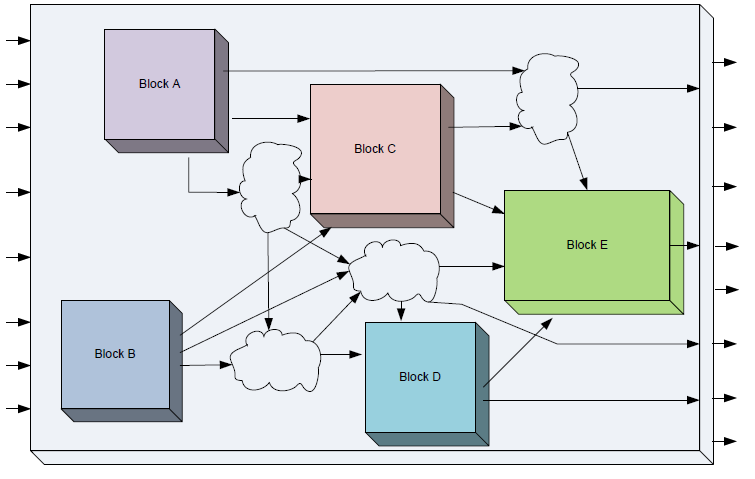}
 \caption{\small Block Level Diagram of a Design}
 \label{bbox}
 \end{center}
 \end{figure}

A representative design for discussion is depicted in
Figure \ref{bbox}. There are 5 blocks \{A-E\} with associated logic
which represent the complete functionality of the design.
\par
When these independent blocks are of decent size, a
complete conglomeration of these blocks would be of a size
which crosses a typical limit of the tool handling capacity. With an
RTL2RTL FEV, the total design size is twice the size of
one design and hence the complexity gets even more
cumbersome. Hence, we resort to first proving the largest
independent blocks or a set of blocks first, for example A
and B blocks separately and then black box them in the
design when we do the top level equivalence.
\subsection{ Selective Enabling and Careful Carving the Logic}

Some of the changes in the design would be very limited to
a set of blocks and hence would not necessarily mandate a
complete block level equivalence. We selectively enable certain blocks and black box the
remaining unwanted logic. As the inputs to those block
boxes would be considered as outputs and the outputs of
those blocks would be considered as inputs, we would need
to have those kinds of tools or methodologies to
automatically map those black boxes and the respective
signals. Careful attention has to be taken while choosing
the logic that has to be black boxed making sure that we
don’t accidentally turn off certain blocks where the changes
might be influential.
\newline
Not all the times, the design changes are limited to one or
two blocks individually, but do span across those blocks
and hence individual equivalence is not always the
preferred solution. Proper care needs to be exercised in carving out the necessary logic for the equivalence.
\subsection{Appropriate Input Constraints or Pruning:}

The design would have certain definitive functionality in
case of a set of valid input constraints. It usually helps if appropriate prunings are applied on the
design so that the input space is constrained and valid.
Though the argument is valid that the design should behave similar
in case of invalid inputs, it does not really add much value to
the validation in question at the cost of increased
complexity.
\subsection{Case splitting:}

Even after pruning the design, the complexity would not
have been completely controlled and needs additional
methods to converge the design. One of the best known
methods is to split the probable inputs into different
subsets. An example for the same sort of case splitting is
discussed in section \ref{bigdata_section}.
\subsection{Helper Assertions:}

One other technique of controlling the complexity is to pick
an intermediate point in both the designs and prove the
equality of such point. Once such a point is identified and
proven, we can use that as a helper assertion and prove the
downstream logic. Effectively the cone of influence (COI)
is subdivided into a simpler cone and a bigger portion of
the cone can converge at a faster pace. More such helper
assertions or equivalence points can help in much faster
convergence.\par

\begin{figure}[H]
 \begin{center}
 \includegraphics [scale = 0.85] {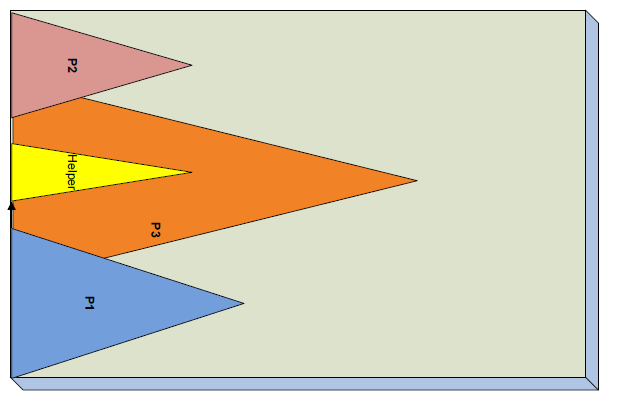}
 \caption{\small Helper Assertions for Faster Convergence}
 \label{helper}
 \end{center}
 \end{figure}

Figure \ref{helper} explains the case where there are certain points
P1, P2 and P3, where P3 is hard to converge, the
equivalence point in both the designs would reduce the
cone for the convergence which is depicted by the helper
cone and that could help in proving P3 faster.
\subsection{State Splitting:}

In cases where it becomes a little bit involved in finding an
exact equivalent point, a routine could be written to
selectively equate certain intermediate points (state
matching) and first prove those. If those intermediate points
fail in the equivalence, remove the mappings of those
points and use those particular points which are proven in
both the designs as the helpers and try to converge to the
final outputs. This kind of state splitting is very helpful in
the cases of timing fixes where certain logic is redistributed
across pipeline stages and the design size is huge and
convergence is an issue.
\newline
\begin{figure}[H]
 \begin{center}
 \includegraphics [scale = 0.85] {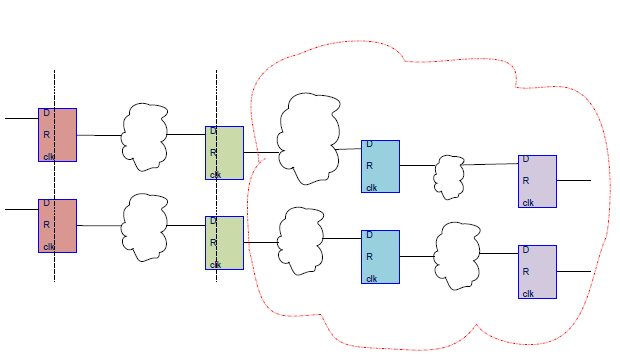}
 \caption{\small Automatic state mapping for faster convergence}
 \label{convergence}
 \end{center}
 \end{figure}
Figure \ref{convergence} depicts one such case of automatic state splitting
where the first two flop stages could be proved first and the
logic shown in the red structure would be allowed to be
proven by the checker. This way the convergence becomes
much faster and cleaner. There are indeed some cases
where we deployed scripts to first map all the flops across
the design between SPEC and IMP and selectively removed
those failing state elements from the mapping and ran it
iteratively. Our modus operandi for this kind of logic is to
first enable the script and check for the equivalence by
brute force method. We also proved the cover
points for those points which should definitely be different
and convince ourselves that we indeed have exercised the
logic. Design knowledge is mandatory to do a diligent
equivalence check in some of these involved cases.
\subsection{Abstracting the design as applicable}

One of the most common form of convergence in the
Bounded Model Checking (BMC) form of FPV is to
abstract some of the complex structures like FIFOs,
Counters, RAMs and memory elements. Some tools do
provide methodologies to easily abstract some of these hard
to crack nuts and help in faster convergence when such
logic could not be completely avoided or black boxed for
proving the equivalence.
\section{Potential Areas of further Application}
\subsection{HLS Model Equivalence:}

HLS is now an established methodology to synthesize the
RTL from a high level language specification like
C/SystemC. Some of the designs don’t start with the
regular high level model to start with, but take some base
from the existing design and recode the same in high level
language and take it through HLS and continue coding in
the high level, down further. Another form of usage
would be to equate the generated RTL against the base with
which the high level coding is started. Though this works
seamlessly for smaller blocks, working on bigger blocks is
still a challenge not yet solved by the current tool set.
\section{Results and Conclusion}

The RTL2RTL formal equivalence verification has been
very successful with its implementation in various flavors
at Intel\textsuperscript{\textregistered}. The beauty of this technique is that it would not
need a bigger validation environment setup or complex
assertions to be coded like the regular FPV. The debug also
would be much simpler as the spec is one of the standard
codes which were verified earlier through different means.
\par
New additions to the code should not change the existing
status of the health of the model and that could be easily
guaranteed by deploying the formal equivalence for every
model release of the design. A great amount of net batch,
memory and human resources can be saved by diligent use
of the methodology proposed.
\par 
There were many cases where the regression time was
drastically cut down using the equivalence verification.
Figure11 depicts a sample of multiple units taken through
FEV against the sanity DV regression, where in the FEV
could guarantee 100\% confidence in the verification, while
the DV failed to give any representative number for the
same. There were many instances where the DV passed but
the FV would catch such kind of corner cases which was not
being exposed in the sanity check in regression. Some of
these would have been found in the weekly regression over
many more thousands of seeds, but all such effort was
clearly saved by deploying FEV for all the cases discussed
in the paper.\par
\begin{figure}[H]
 \begin{center}
 \includegraphics [scale = 0.7] {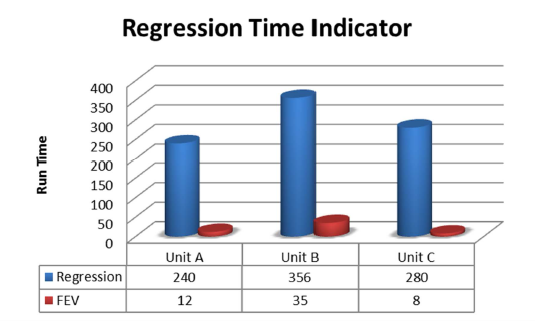}
 \caption{\small Sanity DV regression vs. FEV}
 \label{runtime}
 \end{center}
 \end{figure}

Figure \ref{runtime} shows the time spent on a sanity DV regression
which was effectively saved by deploying an formal
equivalence flow for the design changes while guaranteeing
100\% coverage of all scenarios. Figure 16 talks about one
specific case of STE formal sanity regression replaced with
the RTL2RTL formal equivalence based on an earlier
proven model.\par
\begin{figure}[H]
 \begin{center}
 \includegraphics [scale = 1] {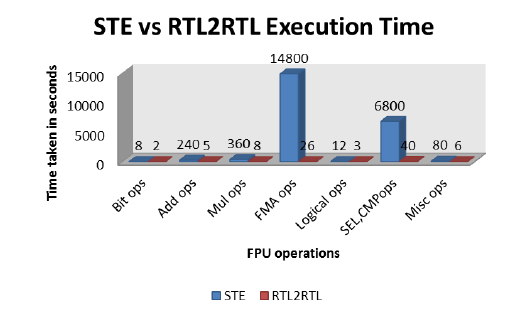}
 \caption{\small Execution time comparison of a complete formal proof vs RTL2RTL equivalence}
 \label{Execution time}
 \end{center}
 \end{figure}

Compared to all the other forms of formal verification, the
equivalence verification is one of the easiest forms and
lowers the barrier for any designer or validator who is not an FV expert or exposed to formal methods. We have deployed several scripts
and methodologies which help the designers use the
RTL2RTL Formal Equivalence as their first line of defense
to release the code to the model.\par

The tools have matured over the time and still striving to
enhance their convergence features. We strongly believe
that many such enhancements would make the adoption of this methodology much easier assisting in faster design convergence.
\paragraph{}
\textbf{ACKNOWLEDGMENTS}
\newline
Sincere thanks to Archana Vijaykumar who has been
supporting us strong in the activity and enabling us to try
out on various designs. We would also like to thank our design team members for the constant
support provided.
\nocite{*}
\bibliographystyle{eptcs}
\bibliography{generic_bindu}
\end{document}